\documentclass[showpacs,showkeys,preprintnumbers,amsmath,amssymb,latexsym]{revtex4}

\usepackage{graphicx}
\usepackage{dcolumn}
\usepackage{bm}

\def\beq{\begin{equation}}
\def\eeq{\end{equation}}

\def\rmd{{\rm d}}

\begin{document}

\title{Strains and axial outflows in the field of a rotating black hole}

\author{D. Bini}
  \email{binid@icra.it}
 \affiliation{Istituto per le Applicazioni del Calcolo ``M. Picone,'' CNR I-00161 Rome, Italy and\\ 
ICRA, University of Rome ``La Sapienza,'' I-00185 Rome, Italy}
\author{F. de Felice}
 \email{fernando.defelice@pd.infn.it}
\affiliation{Dipartimento di Fisica, Universit\`a di Padova, and INFN, Sezione di Padova, Via Marzolo 8,  I-35131 Padova, Italy}
\author{A. Geralico}
  \email{geralico@icra.it}
 \affiliation{Physics Department and
ICRA, University of Rome ``La Sapienza,'' I-00185 Rome, Italy}

\begin{abstract}
We study the behaviour of an initially spherical bunch of accelerated particles
emitted along trajectories parallel to the symmetry axis of a rotating black hole.
We find that, under suitable conditions, curvature and inertial strains compete to model the shape of axial outflows of matter contributing to generate jet-like structures. 
This is of course a purely kinematical effect which does not account by itself for physical processes underlying the formation of jets. 
In our analysis a crucial role is played by a property of the electric and magnetic part of the Weyl tensor to be Lorentz-invariant boosting along the axis of symmetry in Kerr spacetime.
\end{abstract}

\pacs{04.20.Cv}

\keywords{Relative strains, astrophysical jets}

\maketitle

\section{Introduction}

The tidal gravitational accelerations experienced by a steady beam of ultrarelativistic particles rapidly moving with respect to a rotating source along different directions was investigated by  Mashhoon  and McClune \cite{mashmcclune} in the field of a rotating source and in connection with the physics of jets. They considered only gravitational accelerations and argued that the general enhancement of tidal effects could destroy the collimation of the beam and results in the dissipation of the energy of the jet via tidal gravitational radiation, except along special tidal directions as the rotation axis of a collapsed configuration. 

In a number of papers \cite{chicmash1,chicmash2,chicmash3,chicmash4,chicmash5} Chicone and Mashhoon studied the tidal dynamics of relativistic flows near black holes. They investigated the motion of a swarm of free particles (in both non-relativistic 
and relativistic regime) relative to a free reference particle moving on a radial escape 
trajectory. They found that if the relative speed exceeds a certain 
critical value, the first order gravitational tidal effects cause an acceleration/deceleration of the 
ultrarelativistic particles in the swarm moving in directions normal/parallel to the jet direction. 

de Felice and coworkers \cite{fdfcourir,fdfcarlotto} (see also references therein) investigated the conditions for axial collimation of beams emitted near a Kerr black hole. They considered the so called \lq\lq vortical" orbits \cite{fdfcalvani}, i.e. geodesics which spiral around the symmetry axis without crossing the equatorial plane. They found that such a family of particle trajectories collimate towards the rotation axis as a consequence of a constrained variation of their energy and angular momentum, provided the variation was slow enough to preserve the geodesic character of the orbits.

In the present paper we study the relative behaviour of a bunch of particles constrained to move parallel to the axis of symmetry of a Kerr black hole
under the combined effects of a given acceleration and of the background curvature.
Ultrarelativistic particles are expected to be produced near
highly magnetized rapidly rotating neutron stars or as a consequence
of the complicated accretion phenomena in the vicinity
of an active black hole (see e.g. \cite{punsly,guthmann}).
The flow of particles escaping away from the central source and generating polar jets out to
the observed distances appears to be governed by electromagnetic interactions or magnetohydrodynamics.
For our purposes, we imagine an abundance of such particles
near the poles of the collapsed system.
In the case of axial acceleration we find that curvature and inertial strains compete to the formation of jet-like structures from an initial spherical configuration.
The result crucially depends on the property of the curvature to remain finite under ultrarelativistic boost along the axis of symmetry of Kerr spacetime, i.e. approaching a principal null direction of the spacetime itself \cite{maspla1,maspla2,r1,r2,r3,r4}.
The limitations and further developments of this analysis are discussed in the conclusions.
We want to stress that we neither propose a specific model of jets nor make any assumption about the physical processes that are responsible for the initial acceleration, but only discuss in detail a general relativistic kinematical effect that should be considered in any jet model which assumes as a driving machine a rotating black hole.

In this paper Greek indices run from 0 to 3, Latin indices run from 1 to 3 and hatted indices indicate tetrad components. Furthermore units are chosen so that $c=1=G$.

\section{Axial observers in Kerr spacetime}

Consider Kerr spacetime, whose line element in Kerr-Schild coordinates $(t,x^1=x,x^2=y,x^3=z)$ is given by
\begin{eqnarray}
\label{metr_gen}
\rmd s^2 &=& g_{\alpha\beta}\rmd x^\alpha\rmd x^\beta\equiv (\eta_{\alpha\beta}+2Hk_\alpha k_\beta)\rmd x^\alpha\rmd x^\beta\ , \qquad
H=\frac{\mathcal{M}r^3}{r^4+a^2z^2}\ , 
\end{eqnarray}
where $\eta_{\alpha\beta}$ is the flat spacetime metric and
\beq
k_\alpha\rmd x^\alpha=-\rmd t-\frac{(rx+ay)\rmd x+(ry-ax)\rmd y}{r^2+a^2}-\frac{z}{r}\rmd z\ ,
\eeq
with $r$ implicitly defined by
\beq
\frac{x^2+y^2}{r^2+a^2}+\frac{z^2}{r^2}=1\ .
\eeq
Here $\mathcal{M}$ and $a$ are the total mass and specific angular momentum characterizing the spacetime; in geometrized units they have both the dimension of a length. 

Consider then the familiy of static observers on and nearby the symmetry axis. It is understood that these observers only exist outside the ergosphere.
Their four velocity is aligned with the Killing temporal direction
\beq
m=\frac{1}{M}\partial_t\ , \qquad M=\sqrt{-g_{tt}}\ ,
\eeq
with dual
\beq
m^\flat =-M(\rmd t-M_a \rmd x^a)\ , \qquad M_a=-g_{ta}/g_{tt} \quad (a=1,2,3)\ ,
\eeq
where $M$ and $M_a$ are the lapse and shift functions respectively. 
The spacetime metric (\ref{metr_gen}) can then also be written as \cite{mfg}
\beq
\label{metr_thd}
\rmd s^2= -M^2( \rmd t-M_a \rmd x^a )^2 + \gamma_{ab}\rmd x^a \rmd x^b\ ,
\eeq
where 
$$\gamma_{ab}=g_{ab}+M^2M_aM_b= g_{ab}-\frac{g_{ta}g_{tb}}{g_{tt}}\ , \qquad  (a,b=1,2,3)\ .$$

We  now construct an orthonormal frame adapted to the static observers.
First consider the set of three unitary vector fields
\begin{eqnarray}
\label{thdframe}
\epsilon(m)_{\hat 1}=(1/\sqrt{\gamma_{11}})\,(\partial_1+M_1\partial_t)\ , \qquad
\epsilon(m)_{\hat 2}=(1/\sqrt{\gamma_{22}})\,(\partial_2+M_2\partial_t)\ , \qquad
\epsilon(m)_{\hat 3}=(1/\sqrt{\gamma_{33}})\,(\partial_3+M_3\partial_t)\ ; 
\end{eqnarray}
then fix one of these, say $\epsilon(m)_{\hat 3}$, and define the following new vectors
\begin{eqnarray}
f(m)_{\hat 1}=\frac1{\sqrt{1-v_{13}^2}}[\epsilon(m)_{\hat 1}-v_{13}\epsilon(m)_{\hat 3}]\ , \qquad 
f(m)_{\hat 2}=\frac1{\sqrt{1-v_{23}^2}}[\epsilon(m)_{\hat 2}-v_{23}\epsilon(m)_{\hat 3}]\ ,
\end{eqnarray} 
where $v_{ab}=\epsilon(m)_{\hat a}\cdot\epsilon(m)_{\hat b}=\gamma_{ab}/\sqrt{\gamma_{aa}\gamma_{bb}}$; no sum over the indices $a$ and $b$ is meant here.

The unit vectors $f(m)_{\hat 1}$ and $f(m)_{\hat 2}$ are orthogonal to $m$ and $\epsilon(m)_{\hat 3}$ but they are not orthogonal to each other;
however they can be rotated to generate the desired orthonormal frame adapted to the  family of observers $m$: 
\begin{eqnarray}
\label{thrspattriad}
E(m)_{\hat 1}=f(m)_{\hat 1}\ , \qquad 
E(m)_{\hat 2}=\frac1{\sqrt{1-w_{12}^2}}[f(m)_{\hat 2}-w_{12}f(m)_{\hat 1}]\ , \qquad
E(m)_{\hat 3}=\epsilon(m)_{\hat 3}\ ,
\end{eqnarray}
where 
\beq
w_{12}=f(m)_{\hat 1}\cdot f(m)_{\hat 2}=(\gamma_{12}\gamma_{33}-\gamma_{13}\gamma_{23})/\sqrt{(\gamma_{11}\gamma_{33}-\gamma_{13}^2)(\gamma_{22}\gamma_{33}-\gamma_{23}^2)}\ .
\eeq
Note that the frame (\ref{thrspattriad}) is well behaved on the rotation axis; in fact setting $x=0=y$ we have   
\begin{eqnarray}
\label{stat_fr}
m=\sqrt{\frac{z^2+a^2}{\Delta_z}}\partial_t\ , \qquad
E(m)_{\hat 1}=\partial_x\ , \qquad
E(m)_{\hat 2}=\partial_y\ , \qquad
E(m)_{\hat 3}=\sqrt{\frac{\Delta_z}{z^2+a^2}}\left[\partial_z+\frac{2\mathcal{M}z}{\Delta_z}\partial_t\right]\ ,
\end{eqnarray}
where $\Delta_z=z^2-2\mathcal{M}z+a^2=(z-z_+)(z-z_-)$ with $z_\pm = {\mathcal M}\pm \sqrt{{\mathcal M}^2-a^2}$.

Consider now a set of particles moving along the $z$-direction, on and nearby the axis of symmetry, with four velocity 
\beq
\label{boost1}
U=\gamma[m+\nu E(m)_{\hat 3}]\ , \qquad \gamma=(1-\nu^2)^{-1/2}\ ,
\eeq
where the instantaneous linear velocity $\nu=\nu(z)$, relative to the local static observers, is in general a function of the coordinate $z$; here and throughout the paper the physical velocity is in units of light velocity.
These particles are accelerated except perhaps those which move strictly on the axis of symmetry.
Their history forms a congruence ${\mathcal C}_U$ of $\infty^2$ world lines and each of them can be parametrized by the pair $(x,y)$ of the spatial coordinates.
A frame adapted to this kind of orbits can be fixed with the triad
\begin{eqnarray}
\label{radtriad}
E(U)_{\hat 1}=E(m)_{\hat 1}\ , \qquad 
E(U)_{\hat 2}=E(m)_{\hat 2}\ , \qquad
E(U)_{\hat 3}=\gamma[\nu m+E(m)_{\hat 3}]\ ,
\end{eqnarray}
obtained by boosting along $z$ the corresponding triad (\ref{thrspattriad}) adapted to the static observers.

The congruence is accelerated with non-vanishing components along the three spatial directions $E(U)_{\hat a}$ ($a=1,2,3$);
to first order in $x$ and $y$, namely for particles close to the symmetry axis we have
\begin{eqnarray}
a(U)^{\hat 1}= \lambda_1 x+\lambda_2\, y+O(2)\ , \qquad
a(U)^{\hat 2}= \lambda_2\, x-\lambda_1\,y+O(2)\ , \qquad
a(U)^{\hat 3}= a(\tilde U)^{\hat 3}+O(2)\ ,
\end{eqnarray}
where
\begin{eqnarray}
\lambda_1&=& \frac{{\mathcal M}^2}{\Delta_z (z^2+a^2)^{3}}\left[
(5z^2+a^2)\Delta_z+z^4-a^4-\frac{2}{1-\nu}(z^4-a^4-2z^2 \Delta_z)
\right]
\nonumber \\
\lambda_2&=& -\frac{{\mathcal M}z}{\Delta_z (z^2+a^2)^{3}}\left[ 
(5z^2+a^2)\Delta_z+2a^2(z^2+a^2)-\frac{8{\mathcal M}a^2z}{1-\nu}
\right]\ ,
\end{eqnarray}
and $\tilde U$ is the restriction of the vector field $U$ on the axis of symmetry.
It is now possible to study the components of the vector $Y$ connecting the specific curve $\tilde U$ of the congruence ${\mathcal C}_U$, which we fix 
along the rotation axis with $x=0=y$ with tangent vector $\tilde U$, as stated, and nearby  world lines of the same congruence.
The reference world line is accelerated along the direction $E(\tilde U)_{\hat 3}$, i.e. $a(\tilde U)=a(\tilde U)^{\hat 3}E(\tilde U)_{\hat 3}$, with
\begin{eqnarray}
\label{acc}
a(\tilde U)^{\hat 3}=\frac{\gamma}{\sqrt{\Delta_z(z^2+a^2)}}\left[\gamma^2\nu\frac{\rmd \nu}{\rmd z}\Delta_z+ \mathcal{M} \frac{z^2-a^2}{z^2+a^2}\right]
=\frac{\rmd}{\rmd z}\left(\frac{\gamma}{\sqrt{\tilde \gamma{}_{33}}}\right)\ ,
\end{eqnarray}
where $\tilde \gamma{}_{33}=(z^2+a^2)/\Delta_z$ denotes the restriction on the axis of that metric coefficient.
Clearly $a(\tilde U)^{\hat 3}$ has the dimension of the inverse of a length.

\section{Tidal forces, frame induced deformation and strains}

Before proceeding let us remind what is the physical meaning of the connecting vector. 
When expressed in a tetrad frame it gives the spatial separation that would be measured in the given frame between the fiducial observer $\tilde U$ and a nearby particle which belongs to its instantaneous rest space.  As the time goes, this particle may be seen to approach the observer or move away or eventually keep the initial position.
The connecting vector $Y$ is defined to undergo Lie transport along $U$
\beq
\pounds_U Y=0\  \qquad \rightarrow \qquad \nabla_UY=\nabla_YU\ ,
\eeq
leading to the following relevant equations for the components $Y^{\hat 0}$ and $Y^{\hat a}$  along $U$ and with respect to the spatial frame (\ref{radtriad}), respectively:
\begin{eqnarray}
\label{eq0}
&& \dot Y^{\hat 0}=-Y^{\hat a}a(U)_{\hat a}\ ,\\
\label{eq1}
&& \dot Y^{\hat a} +[\omega_{({\rm fw},U,E)}\times \vec Y]^{\hat a} +K(U)^{\hat a}{}_{\hat b}Y^{\hat b}=0\ ,
\end{eqnarray}
or, equivalently, differentiating the latter equation with respect to the proper time of $U$ (hereafter denoted by a dot)
\beq
\label{eq2}
\ddot Y^{\hat a} +{\mathcal K}_{(U,E)}{}^{\hat a}{}_{\hat b}  Y^{\hat b}=0\ ,
\eeq
where 
\beq
\label{Kdef}
{\mathcal K}_{(U,E)}{}^{\hat a}{}_{\hat b}=[T_{({\rm fw},U,E)}-S(U)+{\mathcal E}(U)]^{\hat a}{}_{\hat b}
\eeq
are the components of the \lq\lq deviation'' matrix ${\mathcal K}_{(U,E)}$.

The kinematical tensor $K(U)$ appearing in Eq. (\ref{eq1}) is defined by projecting orthogonally to $U$ through the operator $P(U)^\alpha_\beta=\delta^\alpha_\beta+U^\alpha U_\beta$ the covariant derivative of $U$, i.e. 
\beq
\label{defKdiU}
P(U)_\alpha^\mu P(U)^\beta_\nu \nabla_\mu U^\nu = -K(U)^\beta{}_\alpha\ .
\eeq
Moreover, it is standard to write $K(U)=\omega(U)-\theta(U)$, where 
$\omega(U)$
is an antisymmetric tensor with components $\omega(U)^\beta{}_\alpha$ representing the vorticity of the congruence ${\mathcal C}_U$  and $\theta(U)$  is a symmetric 
tensor with components $\theta(U)^\beta{}_\alpha$ representing the expansion.
$\omega_{({\rm fw},U,E)}$ is a vector representing the angular velocity with which the spatial triad $E_{\hat a}$ rotates with respect to a Fermi-Walker transported triad along $U$
\beq
\label{omegadef}
P(U) \nabla_U E_{\hat a}= \omega_{({\rm fw},U,E)}\times E_{\hat a}\ ,
\eeq
the operation \lq\lq $\times$" denoting exterior product in the observer local rest space; that is, for any $X$ and $Y$ orthogonal to $U$ we have
\beq 
[X\times Y]^a=\epsilon^{\hat a \hat b \hat c}X_{\hat b} Y_{\hat c}\ , 
\eeq
with $\epsilon^{\hat a \hat b \hat c}$ being the Levi-Civita indicator.

The quantity ${\mathcal E}(U)^\alpha{}_\gamma=R^\alpha{}_{\beta\gamma\delta}U^\beta U^\delta$ appearing in Eq. (\ref{Kdef}) is the electric part of the Riemann tensor as measured by the observer $U$.
$S(U)$ is the strain tensor defined by
\beq
S(U)_{\hat a \hat b}=\nabla(U)_{\hat b} a(U)_{\hat a}+a(U)_{\hat a} a(U)_{\hat b}\ ,
\eeq
while the tensor $T_{({\rm fw},U,E)}$ is given by
\begin{eqnarray}
\label{Tdef}
T_{({\rm fw},U,E)}{}^{\hat a}{}_{\hat b}&=&\delta^{\hat a}_{\hat b} \omega_{({\rm fw},U,E)}^2-\omega_{({\rm fw},U,E)}^{\hat a} \omega_{({\rm fw},U,E)}{}_{\hat b} -\epsilon^{\hat a}{}_{\hat b \hat f}\dot \omega_{({\rm fw},U,E)}^{\hat f}\nonumber \\
&& -2\epsilon^{\hat a}{}_{\hat f \hat c}\omega_{({\rm fw},U,E)}^{\hat f} K(U)^{\hat c}{}_{\hat b}\ . 
\end{eqnarray}
We refer to \cite{strains} for a detailed derivation of these equations as well as the discussion about the corresponding observer-dependent analysis.
The study of the relative deviation equation (\ref{eq2}) is essential to have an insight of the interplay between tidal forces, relative strains and properties of the reference triad.

The reference world line plays the role of a \lq\lq fiducial observer" and the most relevant quantities associated with it  are easily obtained by setting $x=0=y$ in their  expressions for the general world line of the congruence.
In order to determine the deviations measured by the \lq\lq fiducial observer" with respect to the chosen  frame
we need to evaluate all the kinematical fields of the congruence (the acceleration $a(U)$, the vorticity $\omega(U)$ and the expansion $\theta(U)$), the electric part of the Weyl tensor (${\mathcal E}(U)$), the strain tensor ($S(U)$) and the characterization of the spatial triad (\ref{radtriad}) with respect to a Fermi-Walker frame  as given by the tensor $T_{({\rm fw},U,E)}$. Their restriction to the reference world line allows for significant  simplifications. To further simplify the notation we shall hereafter omit  the tilde ($\, \tilde {}\, $)  
being understood that all the quantities we shall now consider are defined on the symmetry axis.

Let us proceed evaluating the quantities which enter the relative deviation equation (\ref{eq2}).\\
\noindent
i) The only nonvanishing components of the kinematical tensors $\omega(U)$ and $\theta(U)$ defined by Eq. (\ref{defKdiU}) are given by
\begin{eqnarray}
\label{variousqua_rad}
\theta(U)_{\hat 3 \hat 3}
=\frac{\rmd}{\rmd z}\left(\frac{\gamma\nu}{\sqrt{\gamma_{33}}}\right)\ , \qquad
\omega(U)_{\hat 1 \hat 2}=\gamma(1+\nu)\frac{2a\mathcal{M}z}{\sqrt{\Delta_z}(z^2+a^2)^{3/2}}\ . 
\end{eqnarray}
ii) The vector $\omega_{({\rm fw},U,E)}$ introduced in Eq. (\ref{omegadef}) is given by 
\beq
\omega_{({\rm fw},U,E)}=\omega_{({\rm fw},U,E)}{}_{\hat 3}E(U)_{\hat 3}\ , \qquad 
\omega_{({\rm fw},U,E)}{}_{\hat 3}\equiv\omega(U)_{\hat 1 \hat 2}\ .
\eeq 
iii) The electric part of the Weyl tensor is a diagonal matrix with respect to the adapted frame (\ref{radtriad}), namely
\beq
\label{elermn}
{\mathcal E}(U)=\mathcal{M}z\frac{z^2-3a^2}{(z^2+a^2)^3}{\rm diag}\left[1,1,-2\right]\ .
\eeq
Form (\ref{elermn}) of the electric part of the Weyl tensor coincides with that calculated with respect to the static frame (\ref{stat_fr}), namely ${\mathcal E}(U)_{\hat a \hat b}={\mathcal E}(m)_{\hat a \hat b}$. In fact,
since the fiducial observer moves on the axis of rotation of the Kerr metric and this axis is also a special tidal axis \cite{maspla1,maspla2} then the curvature is unaffected by the Lorentz boost (\ref{boost1}) and (\ref{radtriad}).  
For this reason relation (\ref{elermn}) does not depend explicitly on the relative velocity $\nu$. It is not so however for the inertial terms.\\
\noindent
iv) The only nonvanishing components of the tensor $T_{({\rm fw},U,E)}$ turn out to be 
\begin{eqnarray}
\label{variousqua_rad2}
T_{({\rm fw},U,E)}{}_{\hat 1 \hat 1}&=&T_{({\rm fw},U,E)}{}_{\hat 2 \hat 2}=-\omega_{({\rm fw},U,E)}{}_{\hat 3}^2\ , \nonumber\\
T_{({\rm fw},U,E)}{}_{\hat 1 \hat 2}&=&-T_{({\rm fw},U,E)}{}_{\hat 2 \hat 1}=-\dot\omega_{({\rm fw},U,E)}{}_{\hat 3}\nonumber\\
&=&-\frac{{\mathcal M}a\nu}{(z^2+a^2)^2(1-\nu)}\left[2\gamma^2z\frac{\rmd \nu}{\rmd z}-\frac{(5z^2-a^2)\Delta_z+z^4-a^4}{(z^2+a^2)\Delta_z}\right]\ .
\end{eqnarray}
v) The non-zero components of the strain tensor  $S(U){}^{\hat a}{}_{\hat b}$ can be written as  
\begin{eqnarray}
\label{strainsgen}
S(U){}_{\hat 1 \hat 1}&=&S(U){}_{\hat 2 \hat 2}=-\omega_{({\rm fw},U,E)}{}_{\hat 3}^2+\mathcal{M}z\frac{z^2-3a^2}{(z^2+a^2)^3}\ , \nonumber\\
S(U){}_{\hat 1 \hat 2}&=&-S(U){}_{\hat 2 \hat 1}=-\dot\omega_{({\rm fw},U,E)}{}_{\hat 3}\ , \nonumber\\
S(U){}_{\hat 3 \hat 3}&=&\frac{\gamma}{\sqrt{\gamma_{33}}}\frac{\rmd }{\rmd z}[a(U)^{\hat 3}]+[a(U)^{\hat 3}]^2\ .
\end{eqnarray}
From Eqs. (\ref{elermn}), (\ref{variousqua_rad2}) and (\ref{strainsgen}) it follows that the deviation matrix ${\mathcal K}_{(U,E)}$ defined by Eq. (\ref{Kdef}) has only the non-zero component
\beq
\label{Kab}
{\mathcal K}_{(U,E)}{}_{\hat 3 \hat 3}=-\frac{\gamma}{\sqrt{\gamma_{33}}}\frac{\rmd [a(U)^{\hat 3}]}{\rmd z}-[a(U)^{\hat 3}]^2+{\mathcal E}(U)_{\hat 3 \hat 3}\ ,
\eeq
hence, as expected, the particles emitted nearby the axis in the $z$-direction will be relatively accelerated in the $z$-direction only.
It is straightforward to show that Eq. (\ref{Kab}) can be rewritten as 
\beq
\label{Ktheta}
{\mathcal K}_{(U,E)}{}_{\hat 3 \hat 3}=-\dot\theta(U)_{\hat 3 \hat 3}-\theta(U)_{\hat 3 \hat 3}^2\ .
\eeq

The spatial components of the connecting vector $Y$ are then obtained by integrating the deviation equation (\ref{eq2}) which now reads
\begin{eqnarray}
\label{eq2ordrad}
\ddot Y^{\hat 1}=0\ , \quad 
\ddot Y^{\hat 2}=0\ , \quad
\ddot Y^{\hat 3}=-{\mathcal K}_{(U,E)}{}_{\hat 3 \hat 3}Y^{\hat 3}\ .
\end{eqnarray}
These equations must be solved taking into account also the first order system of Lie transport equations (\ref{eq1}), which become 
\begin{eqnarray}
\label{eq1ordrad}
\dot Y^{\hat 1}=0\ , \quad 
\dot Y^{\hat 2}=0\ , \quad
\dot Y^{\hat 3}=\theta(U)_{\hat 3 \hat 3}Y^{\hat 3}\ .
\end{eqnarray}
Evidently the behaviour of $Y^{\hat 3}$ depends on the sign of $\theta(U)_{\hat 3 \hat 3}$ which can be deduced from Eq. (\ref{variousqua_rad}). 
Differentiating Eqs. (\ref{eq1ordrad}) and using relation (\ref{Ktheta}) one easily gets Eqs. (\ref{eq2ordrad}), as expected.
The system of second order differential equations (\ref{eq2ordrad}) describes the
balancing among tidal forces due to both curvature and inertia so it is essential to interpret the behaviour of the solution $Y^{\hat a}$. 

The components $Y^{\hat a}$ can also be made explicitly 
depending on $z$ instead of the proper time $\tau_U$ using the relation 
\beq
\rmd z/\rmd\tau_U=\gamma\nu/\sqrt{\gamma_{33}}\ ;
\eeq
we find
\begin{eqnarray}
\label{eq1ordradz}
\frac{\rmd Y^{\hat 1}}{\rmd z}=0\ , \quad 
\frac{\rmd Y^{\hat 2}}{\rmd z}=0\ , \quad
\frac{\rmd Y^{\hat 3}}{\rmd z}=\left(\frac{\gamma\nu}{\sqrt{\gamma_{33}}}\right)^{-1}\left[\frac{\rmd}{\rmd z}\left(\frac{\gamma\nu}{\sqrt{\gamma_{33}}}\right)\right]Y^{\hat 3}\ .
\end{eqnarray}
The first two equations imply that the connecting vector components in the $x$ and $y$-directions remain always constant along the path and equal to their initial values.
Last equation of (\ref{eq1ordrad}) can be analytically integrated, so that the solution of the system (\ref{eq1ordrad}) is given by 
\beq
\label{solY}
Y^{\hat 1}=Y^{\hat 1}_0,\qquad Y^{\hat 2}=Y^{\hat 2}_0, \qquad Y^{\hat 3}=C\frac{\gamma\nu}{\sqrt{\gamma_{33}}}\ ,
\eeq
where $C$ is a constant.
This result shows that the conditions imposed on the particles of the bunch to move parallel to the axis of rotation is assured by a suitable balancing 
among the gravitoelectric (curvature) tensor (\ref{elermn}), the inertial tensor (\ref{variousqua_rad2}) and the strain tensor (\ref{strainsgen}).
Were the curvature not invariant under Lorentz boosts, the balance in the transverse direction with respect to the axis of rotation would not be assured.

For any given acceleration $a(U)=A(z)E(U)_{\hat 3}$ of the reference world line  the corresponding instantaneous linear spatial velocity $\bar \nu$, say, can be obtained by integrating Eq. (\ref{acc}):
\beq
\frac{\bar \gamma}{\sqrt{\gamma_{33}}}=\int^z_{z_0} A(z) \rmd z+\bar \kappa \equiv F(z)\ ,
\eeq
implying that 
\beq
\bar \nu=\left[1-\frac{1}{\gamma_{33}F^2(z)}\right]^{1/2}\ , \qquad \bar \gamma=\sqrt{\gamma_{33}}F(z)\ ,
\eeq
where the positive value of the linear velocity has been selected in order to consider outflows. 
The solution (\ref{solY}) for the components of the connecting vector turns out to be
\beq
\label{solYaccgen}
Y^{\hat 1}=Y^{\hat 1}_0\ , \qquad 
Y^{\hat 2}=Y^{\hat 2}_0\ , \qquad 
Y^{\hat 3}=Y^{\hat 3}_0\frac{\bar \gamma\bar \nu}{\bar \kappa \nu_+^0\sqrt{\gamma_{33}}}\ ,
\eeq
where $Y^{\hat a}_0=Y^{\hat a}(z_0)$, $z_0$ is the starting point on the axis and the constant $C$ has been fixed to be $C=Y^{\hat 3}_0/(\bar \kappa \nu_+^0)$, with $\nu_+^0=\nu_+(z_0)$ being the geodesic value (to be defined shortly) of the linear velocity corresponding to outgoing particles evaluated at $z=z_0$.
Geodesic motion along the $z$-axis is indeed characterized by $A(z)=0$ (and so $F(z)=\bar \kappa\equiv E$), implying that
\beq
\label{nugeo}
\nu_\pm=\pm\left[1-\frac{1}{E^2 \gamma{}_{33}}\right]^{1/2}\ , \qquad \gamma_\pm^2 
=E^2 \gamma{}_{33}\ ,
\eeq
where $E=- U\cdot\partial_t$ denotes the (conserved) particle's energy per unit mass.
We notice that timelike geodesics on the axis exist for all values of $z>z_+$ for $E\geq 1$; for $E<1$ the condition for the existence of $\nu_\pm$ is $z_+<z<\bar z$, with
\beq
\label{nucond}
\bar z=\left(\frac{{\mathcal M}}{1-E^2}\right)+\left[\left(\frac{\mathcal{M}}{1-E^2}\right)^2-a^2\right]^{1/2}\ .
\eeq

Let us then consider the case of the reference world line with tangent vector $\tilde U$ constantly accelerated, namely with  $a(\tilde U)^{\hat 3}=A=\,\,$const.
The instantaneous linear velocity relative to a local static observer is given by 
\beq
\label{eq:nua}
\nu_A=\left[1-\frac{1}{\gamma_{33}[\bar\kappa+A(z-z_0)]^2}\right]^{1/2}\ ,
\eeq 
since $F(z)=\bar\kappa+A(z-z_0)$, where the positive value has been selected for $\nu_A$ in order to consider outflows. 
When $A=0$,  $\nu_A\equiv \nu_+$ and $\bar\kappa=E$ acquires the meaning of conserved (Killing) energy of the particle.
The solution (\ref{solY}) for the components of the connecting vector turns out to be
\beq
\label{solYaccrad}
Y^{\hat 1}=Y^{\hat 1}_0\ , \qquad 
Y^{\hat 2}=Y^{\hat 2}_0\ , \qquad 
Y^{\hat 3}=Y^{\hat 3}_0\frac{\gamma_A\nu_A}{\sqrt{\gamma_{33}}\bar\kappa \nu_+^0}\ ,
\eeq
where the constant $C$ of Eq. (\ref{solY}) has been fixed as above with $\nu_+^0=\nu_A(z_0)$.

\begin{figure} 
\typeout{*** EPS figure 1}
\begin{center}
\includegraphics[scale=0.5]{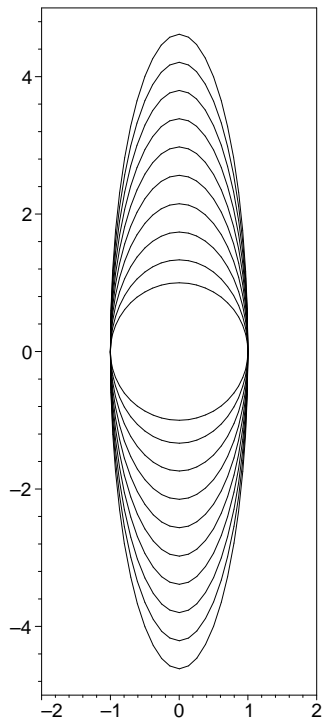}
\end{center}
\caption{The spreading of an initially circular bunch of particles on the $y^{\hat 1}$-$y^{\hat 3}$ plane emitted along the $z$-axis at $z_0/\mathcal{M}=2$ is shown for the choice of parameters $a/\mathcal{M}=0.5$, $\bar \kappa=1.5$ and ${\mathcal M}\, A=0.3$.  
The behaviour is similar for different values of  $\bar \kappa$ and  ${\mathcal M}\, A$.
The curves correspond to increasing values of the coordinate $z/\mathcal{M}=[2,4,6,8,10,12,14,16,18,20]$.
It can be shown that the spreading along the $z$ axis becomes faster and faster for high values of ${\mathcal M}\, A$.
}
\label{fig:1}
\end{figure}

Figure \ref{fig:1} shows the behaviour of an initially spherical bunch of particles in the $Y^{\hat 1}$-$Y^{\hat 3}$ plane for increasing values of the coordinate $z$. We clearly see a 
stretching along the $z$ axis leading to a collimated axial outflow of matter,  clearly suggestive of an  \lq\lq astrophysical jet." 

Note that in this case the acceleration acts contrarily to the curvature tidal effect namely, as we have already seen, $S(U)$ and ${\mathcal E}(U)$ act in competition leading to a quite unexpected result.

Solution (\ref{solYaccgen}) for the components of the deviation vector is formally equal to Eq. (\ref{solYaccrad}), the generality of the situation being implicit in the function $F(z)$.
This suggests that the stretching behaviour shown in Fig. \ref{fig:1} for uniformly accelerated outgoing particles persists also with a general acceleration. This behaviour appears to be independent of the acceleration mechanism itself. This is interesting because accelerations are naturally expected in astrophysical processes.
Observed jets, for instance,  are always associated with magnetohydrodynamical properties which are expected to provide a suitable acceleration mechanism. Whatever the mechanism is,  
this analysis is consistent with their evolution.


\section{Concluding remarks}

We have considered a bunch of test particles moving parallel to the rotation axis of the Kerr black hole spacetime and studied the spatial deviations from a reference world line of the congruence selected to be the one corresponding to a motion along the axis of the hole. In particular we have analyzed a case which can be treated analytically, namely that of the reference world line being  uniformly accelerated.
Deviations from the reference world line there exist in the $z$-direction only. Starting from an initially circular distribution of particles we see  a stretching in the $z$-direction, which becomes faster and faster for increasing values of the acceleration parameter. 
Evidently the result critically depends on the initial conditions and on the constraints imposed on the orbits to be strictly parallel to the axis of symmetry of a rotating black hole. We require an efficient mechanism of confinement which may well be assured by a magnetic field but also by \lq\lq funnels" of massive accretion disks around the axis of rotation whose consideration goes beyond the scope of the paper.

\section*{Acknowledgments}

The authors are indebtet to Prof. B. Mashhoon for many useful discussions. 
DB and AG are grateful to Prof. R. Ruffini and the ICRANet for support.

\end{document}